\documentstyle[prb,aps]{revtex}
\begin{document}
\draft
\title{Phonon Universal Transmission Fluctuations and
Localization in Semiconductor Superlattices with a Controlled
Degree of Order
}
\author{
Norihiko Nishiguchi and Shin-ichiro Tamura \\
}
\address{Department of Engineering Science, Hokkaido University,
Sapporo 060,
Japan \\}
\author{Franco Nori \\}
\address{Department of Physics, The University of Michigan,
Ann Arbor, MI  48109-1120}
\date{\today}
\maketitle
\begin{abstract}
We study both analytically and numerically phonon transmission fluctuations
and localization in partially ordered superlattices with correlations
among neighboring layers.   In order to generate a sequence of layers with
a varying degree of order we employ a model proposed by Hendricks
and Teller as well as partially ordered versions of deterministic
aperiodic superlattices.
\  By changing a parameter measuring the correlation among adjacent layers,
the Hendricks-Teller superlattice exhibits a transition
from periodic ordering,
with alternating layers, to the phase separated
opposite limit; including many
intermediate arrangements and the completely random case.
\ In the partially ordered versions of deterministic superlattices,
there is short-range order (among any $N$ consecutive layers) and long range
disorder, as in the N-state Markov chains.
\ The average and fluctuations in the transmission, the backscattering rate,
and the localization length in these multilayered systems are calculated
based on the superlattice structure factors we derive analytically.  The
standard deviation of the transmission versus the average transmission
lies on a {\it universal \/} curve irrespective of the specific type of
disorder of the SL.
\ We illustrate these general results by applying them to several GaAs-AlAs
superlattices for the proposed experimental observation
of phonon universal transmission fluctuations.
\end{abstract}
\pacs{62.65.+k, 63.50.+x, 68.65.+g, 71.55.Jv}

\narrowtext

\section{Introduction}

For a long time, electronic devices were made of a single semiconductor
material.  This is no longer the case.  Epitaxy and heterostructures have
brought a revolution in device technology by placing different semiconductors,
with different physical properties (dielectric constants, energy gaps, etc.),
within distances of a few nanometers.  For example, different lattice sizes
in different adjacent semiconductors produce strain in the heteroepitaxy,
altering its physical properties.
\ Furthermore, recent developments in the technology for stacking different
semiconductors, in order to fabricate multilayered thin-films, makes possible
the realization of various semiconducting superlattices (SL's) with
artificially imposed one-dimensional (1D) order in the growth
direction.  Specifically, in addition to the usual periodic stacking of
semiconductors, several aperiodic multilayers have been fabricated, including
quasicrystalline, Thue-Morse, and random superlattices.  Their physical
properties have been studied by a variety of experimental probes, including
X-ray and Raman scattering.  For a review on these topics, with further
references, the reader is referred to Ref.\onlinecite{merlin1}.
 We note that the experimental studies of acoustic wave propagations
(of both phonon and ultrasonic regimes )
in some of these aperiodic systems have also been done
by several groups\cite{QCSL,Zhu}.

It is the purpose of this work to study systematically the
  phonon transport properties of
superlattices as a function of their structural order.  In particular,
we study, both analytically and numerically, the transmission fluctuations
and localization properties of phonons in two types of partially ordered SL's,
which are described in more detail in the next two sections.
\ The first type is based on the Hendricks-Teller (HT) model\cite{HT} for
layered systems, which has the very convenient feature of having a tunable
degree of structural correlation among neighboring layers.  In particular, we
consider the gradual and systematic transition from a periodic arrangement
of alternating layers to the opposite, phase separated, regime and follow
the corresponding changes in the transport properties induced by the changing
structural order of the SL.   The second type is based on the so-called
three- and four-state Markov structures\cite{markov1}, and is
illustrated with two examples, which are partially-ordered
versions\cite{merlin_markov} of the quasicrystalline
(QC)\cite{merlin2,QCSL} and Thue-Morse (TM) SL's\cite{TMSL}.
It should be noted that the random version of QC SL's as defined by
a three-state Markov process were also fabricated and
the  Raman spectra in these systems have already been measured\cite{merlin1}.

Our strategy is the following: we derive analytical expressions of $I_s$,
the average phonon intensity reflected from the interface of layers, for
several SL's with a varying degree of short-range correlations.
\ From $I_s$, we analytically derive
the localization lengths, transmission rate, and
transmission fluctuations, all
of which coincide well with the numerical results we obtain from
the alternative
transfer matrix method.  The relation between the different quantities
which characterize phonon transport is presented.  For instance, the Lyapunov
exponent, which provides the inverse of the phonon localization length, is
the logarithmic decrement of the transmission coefficient averaged over
the realizations of disorder.
\ We apply the general ideas and results derived here to several particular
realizations of GaAs-AlAs SL's which are readily accessible
experimentally.  Our predictions for the universal phonon transmission
fluctuations can be tested using currently existing experimental techniques
in phonon spectroscopy
and phonon imaging which have so far been used to verify
the existence of phonon filtering actions of periodic and
QC SL's\cite{QCSL,Nar,Kob,Hur,San}.
The analogies and differences with the universal
conductance fluctuations for transport in disordered systems\cite{ALS}
and speckle phenomena\cite{speckle} will also be discussed.

In sections II and III, we describe in detail the two novel families of
superlattices considered here.
\ In section IV, the phonon backscattering rate is studied in terms of the
structure factors of the SL's, which we derive in closed form.  Section V is
devoted to the average transmission, transmission fluctuations and the
Lyapunov exponent.
\ We illustrate in Sec.VI our general results by applying them
to several proposed GaAs-AlAs superlattices, for the experimental
observation of phonon transmission fluctuations.
\ Section VII presents a summary of our results.

\section{Hendricks-Teller Superlattices}

Consider a SL with two kinds of layers, hereafter denoted by $A$ and $B$,
occurring with frequencies $f_A$ and $f_B$ ($f_A + f_B=1,\ f_A\geq f_B$).   To
introduce a correlation, consider two adjacent layers and denote by $Q_{AA}$
the probability that layer $A$ is followed by layer $A$, $\ Q_{AB}$ the
probability that $A$ is followed by $B$, and so on.  The first layer of the
pair is $A$ ($B$) with probability $f_A$ ($f_B$), thus
\begin{equation}
Q_{AA}+Q_{AB}=f_A,  \hspace*{0.4in} Q_{BA}+Q_{BB}=f_B;
\label{E1}
\end{equation}
\noindent
and similarly
\begin{equation}
Q_{AA}+Q_{BA}=f_A,  \hspace*{0.4in}  Q_{AB}+Q_{BB}=f_B \ .
\label{E2}
\end{equation}
\noindent
{}From Eqs.~(\ref{E1}) and (\ref{E2}) we find
\begin{equation}
Q_{AA}=f_A-1/4+q,  \hspace*{0.3in}   Q_{BB}=f_B-1/4+q,  \hspace*{0.3in}
Q_{AB}=Q_{BA}=1/4-q,
\label{E3}
\end{equation}
\noindent
where $1/4-f_A<q<1/4$, and $q$ measures the degree of correlation among
adjacent layers.  Note that for $f_A=f_B=1/2$, $q=0$ (i.e., no correlation)
corresponds to the completely disordered case.  Let $P_{AB}=Q_{AB}/f_A$ be
the probability that the second layer of the pair is $B$ if layer
$A$ is now introduced as the first layer of the pair.  This conditional
probability can be defined for any pair of layers (e.g., $P_{BA}$). For
convenience, we also introduce $P_{(AA|B)}$ describing the conditional
probability that the $B$ layer is generated after the pair of layers $AA$,
and so on.

In the Hendricks-Teller structure\cite{HT}, the probability that a layer is
present in a certain position depends on the neighboring layers as well as the
abundance of the layer in question.
By changing the value of the parameter $q$
measuring the correlation among neighboring layers, it is possible to
conveniently obtain a variety of different arrangements ranging from the
alternating checkerboard-like periodic pattern to the phase segregated case.
For a positive value of $q$, the same kind of layers tend to stack side by
side; thus, as $q$ increases, the number of the interfaces
between the layers $A$ and $B$ decreases.  For $q=1/4$, the system becomes a
phase separated single hetero-structure where
every layer of material $A$ ($B$)
is attached to material $A$ ($B$), except at the only interface.  The case
$q=0$ corresponds to the completely random SL, which has been studied in
detail in Ref.\onlinecite{nt92}.
For a negative value of $q$, layers $A$ and $B$ tend to stack
in an alternating fashion, 
and as $q$ decreases the system becomes closer to a periodic SL which is
attained for $q=-1/4$ ( with $f_A=f_B=1/2$).
\ In summary, a negative $q$ encourages alternation among layers while a
positive $q$ favors phase segregation.

The study of quasicrystalline diffraction patterns has been partly responsible
for a renewal of interest in the HT model\cite{garg}.  Several variations of
it have been considered.  In one of them, the independent random variables are
the spacings between the planes (or scatterers).  In another one,
the planes are first periodically spaced and then randomly displaced.  This
difference is not physically significant from the point of view of
speckle\cite{speckle,garg}.

\section{N-state Markov superlattices}

Let us now consider a different type of SL with a controlled degree of
randomness.  It is modelled after
the so-called Markov property in the theory of fluctuations, noise,
and stochastic processes\cite{stoch1}.  In fact, the subclass of Markov
systems is by far the most important stochastic process in physics
and chemistry\cite{markov1,stoch2}.
\ Since SL's based on this structure are not well known in the
multilayer community, \  it is worthwhile to explain the origin and motivation
for this kind of system, and a few results useful for calculations in the
next few sections.

The oldest and best known example of a Markov process in physics is
Brownian motion.  If a series of observations of the same Brownian particle
gives a sequence of locations, $\vec{r}_1, \ \vec{r}_2, \ \ldots, \ \vec{r}_n,
\ \vec{r}_{n+1}, \ \ldots \;$, each displacement,
$\vec{\delta}_{n+1}=\vec{r}_{n+1}-\vec{r}_{n}$ is affected by chance, and
its probability distribution only depends on $\vec{r}_{n}$, and
is independent of the previous history
$\vec{r}_{n-1},\ \vec{r}_{n-2}, \ \ldots \ $.
\ Thus, on the sequence of time intervals imposed by a particular
experiment, the position and the velocity of the particle are
Markov processes.  This picture forms the basis of the theory of Brownian
motion.  Other examples of Markov processes
are: the radioactive nuclear decay,
the escape of gas molecules through a small leak,
the destruction of cells by radiation,
and the emission of light by excited atoms.
\ In all of them, on the sequence of time intervals imposed by a
series of measurements, the state of the system at time $t_n$, {\it only \/}
depends on the state at time $t_{n-1}$, and is
independent of the states at all previous times,
$t_{n-2}, \ t_{n-3}, \ \ldots $ .
\ Also, the concept of a Markov process is not restricted to one-component
processes, but applies to $m$ components as well.  The three velocity
components of a Brownian particle and the $m$ chemical components of a
reacting mixture are two examples.

In the previous paragraphs, the word `process' was used in its standard
physics manner, i.e., usually referring to the time evolution of a system.
\ However, sometimes the underlying evolving variable is not time, but space.
\ An example is given by the following Markov process: the loss of cosmic
ray electrons in an absorbing material, the traversed thickness
playing the role usually assigned to time.
\ Here, we also consider space (along the growth direction)
 in a similar manner.
\ Also, in this paper as in most experiments, we build our structures
one layer at a time.  Starting with layer $X$, we add the next layer,
either $X$ or $Y$, according to the probabilities
$Q_{XX}$ and $Q_{XY}$.  Thus, in a Markov SL the addition of any new
layer only depends on the type of layer (or block of layers) most recently
added, and not on the previous ones.

We now proceed to describe Markov 
SL's with short-range correlations in the sequence of constituent layers.
\ Specifically, we consider versions of the quasicrystalline and Thue-Morse
SL's lacking long-range coherence.  The deterministic quasicrystalline
or Fibonacci sequence has long-range order manifested by the presence of
a dense set of Bragg peaks in its structure factor\cite{qc}.
It is generated by
iterating the substitution rules $A \rightarrow AB$ and $B \rightarrow A$,
so only three possible neighboring pairs of layers $AA$, $AB$, and $BA$
(three states) are present, and the $BB$ pair never appears.  In order to
preserve this short-range ordering, 
we generate a Markov sequence based on it, by using the following
straightforward
three-state Markov chain rules: $(i)$ layer $B$ is generated with
probability one after the pair of layers $AA\ $, i.e., $P_{(AA|B)}=1$,
because $AAA$ is a forbidden arrangement in the original system with
long range order; $(ii)$ layer $A$ is generated with probability one
after $AB$, i.e., $P_{(AB|A)}=1$, because $BB$ is not allowed in the original
structure with long-range order; and $(iii)$ $P_{(BA|A)}=\tau^{-1}$  and
$P_{(BA|B)}=\tau^{-2}$, respectively, where $\tau=(\sqrt{5}+1)/2$.
\ The last step is the only one that introduces randomness in this structure.
\ Therefore, the probability of occurrence (i.e., frequency)
for the layers themselves  are $f_{A}=\tau^{-1}$  and $f_{B}=\tau^{-2}$.
 All these probabilities also apply
to the original deterministic structure with long-range order.
\ However, the Markov sequences generated according to the above
probabilistic rules lack long-range coherence.

The Thue-Morse (TM) chain\cite{TMSL,CSM}  is a deterministic sequence that has
a degree of order intermediate between the quasiperiodic and random cases.  In
spite of its aperiodicity, the TM Fourier spectrum exhibits very prominent
peaks that would be absent in a random sequence.  It is the
scaling invariance of the TM chain (periodicity on a logarithmic scale) which
produces long-range correlations.   Many different prescriptions can generate
the TM sequence, the simplest one is through the substitution rules:
$A \rightarrow AB$ and $B \rightarrow BA$.  In this sequence, the adjacent
pairs $AA$, $AB$, $BA$, and $BB$ (four states) appear with equal probability,
and blocks $AAA$ and $BBB$ are not allowed.

We can generate a partially disordered structure, which preserves the TM
short-range arrangement among adjacent layers, by following the simple rules:
$(i)$ layer $B$ $(A)$ is added after layer $AA$ $(BB)$ with probability one,
i.e., $P_{(AA|B)}=1$ and $P_{(BB|A)}=1$;
$(ii)$ add layers $A$ and $B$, with equal probabilities,
after the pairs of layers $AB$ and $BA$, e.g., $P_{(AB|A)}=1/2$ and
$P_{(BA|A)}=1/2$. Note that $f_A=f_B=1/2$ holds in the partially ordered
TM sequences.

\section{ Backscattering Rate of Phonons}

In a recent paper, we have shown that the transmission rate, localization
length and transmission fluctuations of phonons in random SL's are derived
from the backscattering rate of phonons due to mass density fluctuations
in SL's\cite{nt92}.
Incorporating all of the forward scattering contributions,
we can relate the scattering rate to the ensemble average of the squared
SL structure factor, $I_s$, defined by
$I_s= \langle |S_N|^2\rangle /N$, where
\begin{equation}
S_N \ =\ \sum_{j=0}^N  \ (-1)^{j} \ \exp( - i\sum_{m=0}^j \theta_m),
\label{E4}
\end{equation}
is the structure factor of a SL ($\theta_0=0$).
Here, $N$ is the number of $A$- and $B$-blocks
(several consecutive identical $A$-layers define an $A$-block, or block $A$)
in the SL, and $N-1$ is the number of  interfaces between blocks $A$ and
$B$.
\ We use the words ``layer'' and  ``block''  in the following
way: an elementary or basic layer made of material $A$, with thickness $d_A$,
is called an $A$-layer or layer $A$.  Also, $n$ consecutive $A$-layers
form an $A$-block, or block $A$.  The same notation applies to $B$.
\ In Eq.~(\ref{E4}), $\theta_m$ denotes twice the phase factor
which phonons gain in passing through the $m$th block of a SL consisting of
a disordered sequence of layers $A$ and $B$.  More
explicitly, starting from an $A$ block, $\theta_{2j-1}=2k_AD_{2j-1}$ and
$\theta_{2j}=2k_BD_{2j}$,
($\; j=1, 2, \ldots \;$) where $k_A$ and $k_B$ are the wave numbers
of phonons in $A$ and $B$ layers and $D_{2j-1}$ and $D_{2j}$
are the thicknesses of the $(2j-1)$th and $2j$th blocks in a
random SL consisting of $A$ and $B$ layers, respectively.
[~ Note that $D_{2j-1}=(n_A)_{2j-1} d_A$ and $D_{2j} = (n_B)_{2j} d_B$,
where $(n_A)_{2j-1}$
and $(n_B)_{2j}$ are the
 number
of consecutive $A \ (B)$ layers making the $2j-1$th \ ($2j$th)  block,
and $d_A$ and $d_B$ are the thicknesses of the basic or elementary $A$ and $B$
layers. ]
\ The interface between two consecutive identical layers (e.g., the interface
between the $A$ and $A$ layers in a $AA$ block) does not produce any
scattering, therefore, the only relevant interfaces are between
different types of layers or the interfaces between $A$ and $B$ blocks.

Now, we calculate the intensity $I_s$ for the partially ordered SL's.
\ Assuming $N$ is an even number ( $ N = 2 n $),
we can rewrite Eq.~(\ref{E4}) as
\begin{equation}
S_N=\sum_{j=1}^n \left[ 1-\exp( - i\theta_{2j}) \right] \
\exp( - i\sum_{m=1}^{2j-1} \theta_m).
\label{E5}
\end{equation}
Thus, we obtain
\begin{eqnarray}
|S_N|^2  & = & \sum_{j=1}^n \; (1-e^{ - i\theta_{2j}})   \nonumber \\
         &   & - \sum_{j=2}^n  \ \
\sum_{m=1}^{j-1} \ (1-e^{-i\theta_{2j}}) (1-e^{-i\theta_{2m}})
e^{-i ( \theta_{2j-1} + \theta_{2j-2} + \ldots + \theta_{2m+1} ) } + c.c.
\label{E6}
\end{eqnarray}
We consider the case where no correlation exists between the thicknesses of
the adjacent blocks 
(this is valid for the case we are considering) and put
\begin{eqnarray}
\langle \exp( - i\theta_{2j-1})\rangle & = &
\langle \exp( - 2 i k_A D_{2j-1})\rangle
\ \equiv \epsilon_A,
\nonumber \\
\langle \exp( - i\theta_{2j})\rangle   & = &
\langle \exp( - 2 i k_B D_{2j})\rangle
\ \equiv \epsilon_B.
\label{E7}
\end{eqnarray}
Now, it is straightforward to derive the expression of $I_s$ for
$ |\epsilon_A \epsilon_B|<1$\cite{tn90}.  The result is
\begin{equation}
I_s= Re \left[ \frac{(1-\epsilon_A)(1-\epsilon_B)}
{1- \epsilon_A \epsilon_B} \right] \; .
\label{E8}
\end{equation}
To proceed further, we calculate $\epsilon_A$ and $\epsilon_B$ for any given
partially-random sequence of $A$ and $B$ layers.
 For the Hendricks-Teller model,
we find the following averaged phase factors
\begin{equation}
\epsilon_A = \frac{P_{AB}\ e^{ - ia}}{1-P_{AA}\ e^{ - ia}}\ , \hspace*{0.6in}
\epsilon_B = \frac{P_{BA}\ e^{ - ib}}{1-P_{BB}\ e^{ - ib}}\ ,
\label{E9}
\end{equation}
and averaged block lengths
\begin{equation}
\langle D_{2j-1}\rangle  \equiv  \langle D_A\rangle
=\frac{P_{AB}}{(1-P_{AA})^2}
\; d_A,  \hspace*{0.4in}
\langle D_{2j}\rangle \equiv  \langle D_B\rangle =
\frac{P_{BA}}{(1-P_{BB})^2} \; d_B \;,
\label{E10}
\end{equation}
where $a=2k_A d_A$ and $b=2k_B d_B$.
Similarly, for the three-state Markov SL with short-range quasicrystalline
order we obtain
\begin{equation}
\epsilon_A = \frac{1}{\tau}e^{ - 2ia} + \frac{1}{\tau^2}e^{ - ia},
\hspace*{0.6in}
\epsilon_B = e^{ - ib}
\label{E11}
\end{equation}
and $\langle D_A\rangle =\tau d_A$ and $\langle D_B\rangle =d_B$.
For the Markov TM SL
\begin{equation}
\epsilon_A = \frac{1}{2}(e^{ - ia} + e^{ - 2ia}),  \hspace*{0.8in}
\epsilon_B = \frac{1}{2}(e^{ - ib} + e^{ - 2ib}),
\label{E12}
\end{equation}
and $\langle D_A\rangle =3d_A/2$ and $\langle D_B\rangle =3d_B/2$.

The explicit expression of $I_s \equiv I_s^{HT}$ for SL's based on HT model
(with $f_A = f_B = 1/2$) is
\begin{equation}
  I_s^{HT} = \frac{2(1+4q)(\cos \phi - \cos \delta )^2}
   {(1+4q)^2 (\cos \phi - \cos \delta )^2 + ( 1-4q)^2 \sin^2 \phi},
\label{E13}
\end{equation}
where $\phi = (a+b)/2$ and $\delta = (a-b)/2$. Also we find
$I_s \equiv I_s^{M-QC}$
and $I_s \equiv I_s^{M-TM}$ for Markov QC and Markov TM SL's as
\begin{equation}
  I_s^{M-QC} = \frac{(1-\cos a)(1- \cos b )}
   {\cos a + \tau \left[ 2- \cos (a+b) \right] - \tau^2 \cos (2a+b)},
\label{E14}
\end{equation}

\begin{equation}
  I_s^{M-TM} =
 \frac{\frac12(1-\cos a)(1- \cos b )(\cos a + \cos b + 5/2)}
   {1+\cos^2 \frac a2 \cos^2 \frac b2
        - \frac 12 \left[ \cos (a+b) + \cos (2a+b) + \cos (a+2b) +
   \cos (2a+2b)\right]}.
\label{E15}
\end{equation}

Now, according to our previous work\cite{nt92}, the elastic backscattering
rate of phonons due to mass-density fluctuations in random SL's is given
in the Born approximation as
\begin{equation}
\Gamma(\omega) \;=\; \frac{c_A}{D_0} \; R^2 \; I_s \ ,
\label{E16}
\end{equation}
where
$\; D_0=(\langle \ D_A\ \rangle +\langle \ D_B\ \rangle )/2=L/N \ $
is the average thickness of one block
in the system ($L$ being the total length of the SL),
and $R =(Z_A - Z_B)/(Z_A + Z_B)$ with
$Z_i = \rho_i c_i$ ( $i = A \, {\rm and }\,
 B$,
and $c_i$ is the sound velocity) is the amplitude reflection coefficient.
Here, we note that both the  substrate and  detector are assumed to be made
of $A$ material and the homogeneous system consisting of only $A$
material is taken as the unperturbed system based on which the calculation
of the phonon transmission rate is developed.
Thus, $D_0/c_A$ gives the average time for phonons to
propagate through the length of
a single block in the unperturbed structure.

\section{ Phonon Transmission Rate, Lyapunov exponent
and Transmission Fluctuations}

In Ref.\onlinecite{nt92}
we have derived a formula which relates the phonon
backscattering rate to the transmission rate.
Introducing a scaling parameter
$t=(L/c_A)\Gamma=L/\ell$ ( $\ell=c_A/\Gamma$ is the elastic mean
free path of
backscattering),
the average transmission rate $\langle T \rangle$ is given by
\begin{equation}
\langle T \rangle=\int^{\infty}_0 \; \
\frac{2\pi \lambda \tanh \pi \lambda}{\cosh \pi \lambda}
\exp\left[-(\frac{1}{4}+\lambda^2)t\right] d\lambda \ .
\label{E17}
\end{equation}
This formula was originally derived in the study of the electrical
conductivity in one-dimensional disordered metals\cite{abrikosov}.

The Lyapunov exponent $\gamma$, defined by\cite{lfs}
\begin{equation}
\gamma = - \lim_{L \rightarrow \infty} \ \langle \log T\rangle /2L,
\label{E18}
\end{equation}
is an important quantity which provides the phonon localization
length $\xi=\gamma^{-1}$.  The Lyapunov exponent is the logarithmic decrement
of the transmission coefficient averaged over the realizations of disorder.
\ It can be proved that $\gamma=\Gamma/(2c_A)=1/2\ell$,
so $\gamma$ is directly related to the structure factor or
$I_s$\cite{nt92}.  Also, the standard deviation of the transmission
\begin{equation}
\Delta T=(\langle T^2 \rangle-\langle T \rangle^2)^{1/2}
\label{E19}
\end{equation}
is calculated from Eq.~(\ref{E14}) by employing the relation \cite{abrssc}
\begin{equation}
\langle T^2 \rangle \ = \ - \ \frac{d\langle T \rangle}{dt} \ .
\label{E20}
\end{equation}

In the next section, we will present comparisons between our analytical and
numerical results for a variety of SL's.  It is important to point out that
the plots presented below are not fingerprints (or speckle patterns) of
specific configurations of
disorder but averages over many realizations of disorder.  The term speckle
pattern refers to the complex interference
pattern in the transmitted intensity
as a function of frequency (or the outgoing direction).  Each
realization of a random medium ({\it i.e.}, each sample of the statistical
ensemble) displays its own pattern, or ``fingerprint'', which reflects the
specific arrangement of the inhomogeneities ({\it e.g.}, impurities) in that
sample.
\ This phenomenon, called ``speckle pattern'', is familiar
in optics\cite{speckle} and
it refers to the intensity pattern formed on a screen by light reflected
from a rough surface.  The detailed study, with experimental predictions,
of the phonon spectroscopy analog of these ``fingerprints''
will be presented elsewhere.

The expression `universal transmission fluctuations'
clearly does not refer to the phonon analog
of `universal conductance fluctuations', but to the fact
that different realizations of disorder have fluctuations which fall on the
very same {\it universal\/} curve for the standard deviation versus average
transmission.
\  In fact, we do obtain, analytically and
 numerically (for a variety of SL's),
 a universal curve  ($ \Delta T$ versus $\langle T \rangle $) for the
 transmission fluctuations.
\ Also, universal conductance fluctuations are not directly related
to localization, while our focus here is on localization.
\ Finally, it has been pointed out that the notation ``universal
conductance fluctuations'' is a misnomer because it refers to a
sample-dependent, and therefore non-universal, fingerprint.  Currently,
they are more appropriately denoted by the term ``reproducible conductance
fluctuations.''

\section{Comparison between Analytical and Numerical Results}
\subsection{Hendricks-Teller model}
\ Figures 1-5 present calculations for the
 quantities described above, obtained
by using two very different approaches.  In one of them, we use the analytical
expressions presented in this paper.
In the other one, we use
the transfer matrix method for numerical calculations.
In the latter method, the displacement and stress fields associated
with the incident and transmitted waves are connected each other
by the product of the transfer matrices describing the physical
properties of each constituent layer of the SL.
The transmission rate is expressed in terms of elements
of the product of the transfer matrices, by imposing proper
boundary conditions on the incoming and outgoing waves.
Readers interested in a pedagogical introduction to
transfer matrices and other related techniques,
are referred to Ref.\onlinecite{lfs}.

 In order to verify the
accuracy of our
predictions, it is important to compare the results obtained from these two
quite different approaches.

Figures 1(a) to 1(c) plot $\langle T \rangle$ versus frequency
 for the Hendricks-Teller
model with $f_A=f_B=1/2$ and for $q=1/8,\ 0$, and $-1/8$.
\ For $q=1/8$, the same kind of layers tend to stack side by
side; $q=0$ corresponds to the completely random SL;
and for $q=-1/8$, layers $A$ and $B$ tend to stack alternatively.
\ For $q=1/4$ the system is phase segregated with a single hetero-structure
and the transmission rate becomes a constant independent of the phonon
frequency.
\ For $q=-1/4$ the system is a periodic SL and sharp dips in transmission
occur due to the Bragg reflection of phonons.
\ Figures 1(a) to 1(c) properly reflect the features characteristic of these
SL systems with a highly-controlled degree of disorder.
\ To plot these
figures we have chosen $34$-\AA-thick GaAs and AlAs as the $A$ and $B$
layers, respectively.  The average transmission reveals various structures
including sharp enhancements and dips.
The former, with $\langle T \rangle \simeq 1$,
are the resonances which occur for phonons whose wavelengths match the
thicknesses of $A$ and $B$ layers. More explicitly, the resonances occur for
$\cos a=1$ and $\cos b=1$, or equivalently at
$\nu=\nu_{i,n}^{(R)}=nc_i/2d_i$ ($i=A$ or $B$, and $n$ is an integer).
\ This can be seen from Eq.~(\ref{E13}) by noting that
$\ (\cos \phi - \cos \delta)^2=
(1-\cos a)(1-\cos b)$.
Numerically, the resonance frequencies are
$\nu_{A,n}^{(R)} = 490 \times n$ GHz
and $\nu_{B,n}^{(R)} = 582 \times n$ GHz.

\ The dips in $\langle T \rangle$ are due to {\it constructive interference\/}
of backscattered
phonons.
\ In the HT SL's the minima of $\langle T \rangle$ are realized at the
frequencies $\nu=\nu_n^{(B)} \equiv n/2 ( d_A/c_A + d_B/c_B)$. These are
the Bragg frequencies ( numerically $\nu_n^{(B)}= 266\times n$ GHz)
in the periodic SL's consisting of an alternating stacking
of $A$ and $B$ layers.
It should noted that $\langle T \rangle$ is monotonically decreasing  as $t$
($\propto \Gamma \sim I_s$) increases, and $I_s$ takes its maximum value for
$\sin \phi = 0$ or $\nu = \nu_n^{(B)}$ (see Eq.~(\ref{E13})).
We find that the overall agreement between the analytical
and numerical results is excellent, even though we observe large fluctuations
in $\langle T \rangle$.  These fluctuations are small only for
$T$ close to zero
and unity,  and remain large even if we increase the system size, i.e.,
the phonon transmission is not a self-averaging quantity, as described
in Refs.\onlinecite{abrssc,andrsn}.

The frequency dependence of the Lyapunov exponent, which is proportional
to $I_s$, is also plotted in Fig.2 for $q=1/8,\ 0$, and $-1/8\ $.  At the
resonances, $\gamma$ vanishes because $I_s=0$ and the phonons are delocalized.
\ The maximum values $\gamma_{max}$ of $\gamma$ are achieved at the
Bragg frequencies $\nu_n^{(B)}$.
\ At these frequencies $I_s=4/(1+4q)$ because
$\sin \phi = 0$ and from Eq.~(\ref{E10})
\begin{equation}
D_0=\frac{2}{1-4q} ( f_A d_A + f_B d_B),
\label{E21}
\end{equation}
so $\gamma_{max}$ are proportional to $(1-4q)/(1+4q)$. This explains the
relative magnitudes of $\gamma_{max}$  for different values of $q$ shown
in Fig.~2.  We also note that in the present case, the localization
lengths are longer than $3000\;$\AA, which is much larger than the typical
wavelength of $35{\rm \AA}$ at a 1-THz frequency.  Even under this condition,
phonons in an infinite, partially ordered SL are localized except at
resonance frequencies due to the coherent interference of backscattered
waves\cite{Ishii}.

Figure 3 exhibits $\Delta T$ versus $\langle T \rangle$ plotted together for
different values of $q$.  It is important to emphasize that the data lies on a
{\it universal \/} curve irrespective of the value of $q$ (magnitude of the
correlation) and also of the specific type of ordering of the SL.  There is
a certain amount of fluctuation in the data numerically computed via transfer
matrices (shown as scattered points in the
 figure).  This fluctuation, however,
decreases when the average is
taken over an ensemble consisting of a larger number of 
SL's.  We have also explicitly shown (see Fig.3)
that the standard deviation $\Delta T$ vanishes for
$\langle T \rangle=1$ and $0$,
monotonically increases in the range
$ \ \langle T \rangle \, \protect\lesssim \, 0.4$,
and monotonically decreases in the range
$  \langle T \rangle  \protect\gtrsim \ 0.4$.

\subsection{QC and TM Markov superlattices}

The average transmission rate versus frequency for the Markov versions
of the quasicrystalline and TM SL's are plotted in Fig.4 together with the
transmission rate of the original QC and TM SL's with long-range deterministic
order.  The basic layers $A$ and $B$ assumed here, are the same ones used for
the HT model.  In the regular QC SL considered here,
the transmission dips occur
at frequencies $\nu_{m,n}\equiv(m+n\tau)v/(2\tau^2 d)$, where $m$ and $n$ are
integers, $v$ is the average sound velocity in the SL, and $d=\tau d_A +d_B$.
\ When $m$ and $n$ are neighboring Fibonacci numbers, i.e.,
$(m,n)=(F_{p-1},F_p)$ where $F_{p+1}=F_p+F_{p-1}$ and $(F_0,F_1)=(0,1)$,
$\nu_{m,n}=\nu_p \equiv \tau^{p-2}v/2d$ holds and a major dip is realized.
\ We have indicated in Fig.4(a) the set of integers $(m,n)$ for several
major dips.  In the regular TM SL we study, large dips in  transmission
happen at frequencies $\nu_n$ and $\nu_{n/3}=\nu_n/3$, where
$\nu_n=\nu_n^{(B)}$.  We have labeled in Fig.4(b) the indices of the
frequencies for several major dips.  We see that the small dips exhibiting
the self-similar structures characteristic of the QC SL\cite{QCSL} are smeared
out in the three-state QC Markov SL, producing rather broad transmission
dips.  Similar results can be seen for the four-state TM Markov case.

Figures 5-7 show the results for the three-state QC and four-state TM Markov
SL's corresponding to Figs.1-3 of H-T model.
\ In Fig.5, the average transmission rate, $\langle T \rangle$,
versus frequency,
$\nu$, is presented and the agreement between the analytically derived
results and the numerical calculations is excellent. \ The transmission
fluctuations are large for an intermediate value of $\langle T \rangle$
as in the case of the HT model.
\ The Lyapunov exponent, $\gamma$, versus frequency  is
 presented in Fig.6.
\ The explicit expressions of $I_s  \, ( \propto  \gamma)$ for the QC Markov
and TM Markov SL's are given by Eqs.~(\ref{E14}) and (\ref{E15}).
\ From these equations we see
that the resonances $(\gamma = 0)$ in these systems occur at the same
frequencies $\nu_{i,n}^{(R)}$ ( satisfying $\cos a = 1$ or $\cos b=1$) as in
the HT SL's. \ Unfortunately, however, we could not find any simple
explicit analytical expression
for the frequencies at which the maximum values of $\gamma$ are attained.
\ The standard deviation of the phonon transmission rate $\Delta T$, versus
$\langle T \rangle$ is shown in Fig.7.
\ In this figure the continuous line is the
theoretical prediction, which agrees well with the numerical points obtained
by averaging over $100$ realizations of disorder.
\ Here, it should be noted that the analytical results for both
$\langle T \rangle$ and  $\Delta T$ are functions of only the scaling
parameter $t=L/\ell$, the  system size divided by the mean-free path.
Thus, $\Delta T$  versus
$\langle T \rangle$  does not depend on the structures of the SL's and the
analytical curve in  Fig.~7 is identical to that in Fig.3 for the HT model.

\section{Conclusions}

In summary, we have derived analytical
expressions of $I_s$, the average phonon
intensity reflected from the interface layers, for several SL's with
varying degrees of short-range correlation.
\ From $I_s$, we derive 
the localization lengths, transmission rate, and transmission fluctuations,
all of which coincide well with the numerical results obtained from the
transfer matrix method.

In the SL's based on the HT model, the introduction of correlations among
neighboring layers drastically changes the
behavior of the phonon transmission.
\ In particular, the rate of transmitted phonons decreases significantly
with decreasing $q$.  Also, the fluctuations in the average transmission
are very small close to  $\langle T \rangle \sim 1$ and
$\langle T \rangle\sim 0$,
and become much larger for intermediate values of $\langle T \rangle$.

The introduction of disorder in the QC and TM SL's produces a
decrease in the phonon long-range coherence which is reflected in the smearing
 out of small peaks  in $I_s$, and equivalently the smearing out of the
small dips  in the  transmission rate.
In spite of these quantitative differences
 in the fine  structure, the overall qualitative behavior is
still the same as in
the ordered case, in the sense that it still exhibits pronounced peaks and
dips in approximately the same locations as in
the original deterministic SL's.
Here we note, however,
that the structure factors in the original, ordered QC and TM SL's have very
sharp peaks
at $\nu=\nu_p$ and $\nu=\nu_{n/3}$, respectively, which means that $I_s$
grows in proportion to the system size or $N$ at these frequencies.  In the
Markov SL's, which only preserve short-range QC or TM ordering,
$I_s$ remains finite on the entire frequency range of phonons,
even if the system size is increased indefinitely.

We have obtained $\Delta T$ versus $\langle T \rangle$ through two different
approaches.   The data lies on a {\it universal \/} curve
irrespective of the value of $q$ (magnitude of the correlation) and
of the specific type of ordering of the SL as demonstrated numerically
for both HT SL's and two kinds of Markov SL's.
\ This is because both $\langle T \rangle$ and
$\Delta T$ are determined only by the magnitude of the elastic mean-free-path
and the system size but does not depend explicitly on the details of the
structure of random SL's.

One of the main findings of this work is that the loss of long-range
order does not produce large qualitative changes in the overall
structure of the phonon transmission rate, while it drastically
affects its fine structure.
\ At first sight, it might seem surprising to see that the very-short-range
correlations among neighboring layers dominate the overall
frequency dependent transmission rate of the traveling phonons.
\ By increasing the degree of ordering in the SL in a controlled manner,
we find that the long-range ordering is responsible for the fine structure
present in the transmission rate.
\ Furthermore, this effect has been precisely quantified through the
computation of Lyapunov exponents and other quantities useful for
describing the localized character of the phonons.  Moreover,
we have applied these ideas to two families of SL's where the degree
of order can be systematically changed in a convenient manner.
\ Finally, we have applied the results
derived here to several particular realizations of GaAs-AlAs SL's
which are readily accessible experimentally.  
\ Thus, our predictions for the phonon universal
transmission fluctuations can be tested using currently existing
experimental techniques in phonon spectroscopy.

Finally, we note that a very interesting experiment has recently
been done by Kono et al, which studies the localization properties of the
third sound waves by directly measuring the transmission spectra
in 1D random lattices\cite{Kono}.
The observed averaged transmissivity reveals the frequency dependence
very similar to those given in Figs. 1(a) to 1(c), i.e.,
the periodic oscillation characteristic of the presence of
both resonances (enhancements in transmission) and
 localizations (dips in transmission ) of the waves.

\acknowledgments

The authors would like to thank T. Sakuma and R. Richardson
for useful comments on the manuscript.
This work is supported in part by the Special Grant-in-Aid for
Promotion of Education and Science in Hokkaido University
Provided by the Ministry of Education, Science and Culture of
Japan,  a Grant-in-Aid for Scientific Research from the Ministry
of Education, Science, and Culture of Japan
(Grant No. 03650001 and No. 05650048),
the Suhara Memorial Foundation,
and the Iketani Science and Technology Foundation.
FN acknowledges partial
support from the
NSF grant DMR-90-01502, a GE Fellowship, and SUN microsystems.

\newpage
\begin{figure}
\caption{ 
Average phonon transmission rate, $\langle T \rangle$,
versus frequency, $\nu$,
for a Hendricks-Teller SL with $200$ layers, $f_A=f_B=1/2$, for
(a) $q=1/8$, (b) $q=0$, and (c) $q=-1/8$.
\ These and the following figures consider $34$-\AA--thick GaAs \ $ (A)$ and
AlAs\ $(B)$ layers as the basic units for constructing the SL's.
The continuous line is our analytical result while the open circles are
 obtained from a numerical calculation using transfer matrices and averaging
over $100$ realizations of disorder.
\ The fluctuations in the average transmission are very small when
$\langle T \rangle\sim 1$,
 and become much larger for the intermediate values of
$\langle T \rangle$, {\it i.e.} for
$0.2 \, \protect\lesssim  \, \langle T \rangle \,
\protect\lesssim  \, 0.6 \;$.
\ Note also that the amount of transmitted phonons decreases significantly
with decreasing $q$.  This can be understood as follows:
\ Similar kinds of layers tend to stack together when $q=1/8$, thus
reducing the number of interfaces with an acoustic mismatch, which are
the scatterers.  For $q=-1/8$, layers $A$ and $B$ tend to
stack alternatively
increasing the number of interfaces (i.e., scatterers).}
\end{figure}

\begin{figure}
\caption{
Lyapunov exponent  $\gamma$  (the inverse of the localization
length $\xi$) versus phonon frequency $\nu$, \, for HT SL's with $q=1/8 \ $
(thin solid line),
$\ 0\ $ (dashed line), and $-1/8$ (bold solid line).
At the resonance frequencies $\nu_{i,n}^{(R)}$, $\ \gamma$ vanishes, and at
the Bragg frequencies $\nu_n^{(B)}$, $\gamma$ takes its maximum value.
The localization lengths are about two orders of magnitude
larger than the typical wavelength of $35$\AA \ at a $1$-THz frequency.
\ Note that the larger the number of interfaces (i.e., for $q=-1/8$),
the sharper the peaks in the Lyapunov exponent as described in the text.
 }
\end{figure}

\begin{figure}
\protect\caption{
Standard deviation of the phonon transmission rate,
$\Delta T=(\langle T^2 \rangle-\langle T \rangle^2)^{1/2}$, versus
$\langle T \rangle$, for the same HT
SL as in the previous figures.  The continuous line is the theoretical
prediction, obtained from Eqs.~(\protect\ref{E17})
and (\protect\ref{E20}).
The points are computed
using the transfer matrix method and an ensemble average over $N_{av}=100$
realizations of disorder.  Fluctuations in the points diminish for increasing
values of $N_{av}$.
}
\end{figure}

\begin{figure}
\caption{
Average transmission rate  $\langle T \rangle$,  versus frequency  $\nu$ for
 the Markov versions (bold solid lines)
of the (a) QC and (b) TM SL's.
Also plotted by thin solid lines are the transmission rates for the original,
deterministic (a) QC and (b) TM SL's with 55 and 64 layers, respectively.
The
physical parameters assumed here are the same ones used for the HT SL.
\ Several major dips in $\langle T \rangle$ are labeled (see the text).
}
\end{figure}

\begin{figure}
\caption{
Average transmission rate $\langle T \rangle$,  versus frequency
$\nu$ of (a) QC and (b) TM Markov SL's.  The continuous
lines are the analytical calculations and the open circles are the
numerical results for the
(a) three-state QC and (b) four-state TM Markov SL's.
The results shown are those averaged over an ensemble of 100 random SL's.
Each random SL consists of 200 basic $A$ and $B$ layers.
}
\end{figure}

\begin{figure}
\caption{
Analytically calculated Lyapunov exponent,
$\gamma$, versus frequency for the QC Markov
(solid line) and TM Markov (dashed line) SL's.
 }
\end{figure}

\begin{figure}
\caption{
Standard deviation of the phonon transmission rate  $\Delta T$
versus average transmission $\langle T \rangle$.
\ The continuous line is the theoretical prediction.
\ Open circles and  squares are the numerical results for the Markov versions
of the QC and TM SL's obtained from the transfer matrix method by averaging
$100$ realizations of disorder.
}
\end{figure}


\begin{references}
\bibitem{merlin1}
R. Merlin, {\it IEEE J. Quantum Electron\/} {\bf QE-24}, 1791 (1988).
\bibitem{QCSL}
S. Tamura and J.P. Wolfe, Phys. Rev. B {\bf 36}, 3491 (1987);
D.C. Hurley, S. Tamura, J.P. Wolfe, K. Ploog  and J. Nagle, {\it ibid}
{\bf 37}, 8829 (1988).
\bibitem{Zhu} Y. Zhu, N. Ming and W. Jiang,  Phys. Rev. B {\bf 40},
8536 (1989).
\bibitem{HT}
S. Hendricks and E. Teller, J. Chem. Phys. {\bf 10}, 147 (1942).
\bibitem{markov1} D. Isaacson and R. Madsen, {\it Markov Chains: Theory and
Applications} (Wiley, New York, 1976).
\bibitem{merlin_markov}
R. Merlin, K. Bajema, J. Nagle  and K. Ploog,
J. Phys. (Paris) Colloq. {\bf 48}, C5-503 (1987).
\bibitem{merlin2} R. Merlin, K. Bajema, R. Clarke, F. Y. Juang and
P. K. Bhattacharya,  Phys. Rev. Lett. {\bf 55},1768(1985).
\bibitem{TMSL} S. Tamura and F. Nori,  Phys. Rev. B {\bf 38}, 5610 (1988).
\bibitem{Nar} V. Narayanamurti, H. L. St\"ormer, M. A. Chin,
A. C. Gossard  and W. Wiegmann, Phys. Rev. Lett. {\bf 43},
2012 (1979).
\bibitem{Kob} O. Koblinger, J. Mebert, E. Dittrich,
 S. D\"ottinger, W. Eisenmenger, P. V. Santos
and L. Ley, Phys. Rev. B {\bf 35}, 9372 (1987).
\bibitem{Hur} D. C. Hurley, S. Tamura, J. P. Wolfe
and H. Morkoc, Phys. Rev. Lett. {\bf 58}, 2446 (1987);
S. Tamura, D. C. Hurley, and J. P. Wolfe,
Phys. Rev. B{\bf 38}, 1427 (1988).
\bibitem{San} P. V. Santos, J. Merbert, O. Koblinger
and L. Ley, Phys. Rev. B {\bf 36}, 1306 (1987).
\bibitem{ALS}
S. Washburn and R.A. Webb, Adv. Phys. {\bf 35}, 375 (1986);
A.G. Aronov and Yu.V. Sharvin, Rev. Mod. Phys. {\bf 59}, 755 (1987).
\bibitem{speckle}
J.W. Goodman, in {\it Laser Speckle and related Phenomena},
edited by J.C.Dainty (Springer-Verlag, Berlin, 1984), 2nd ed.
\bibitem{nt92}
N. Nishiguchi and S. Tamura,
in {\it Proc. Phonon Scattering in Condensed Matter} {\bf VII},
edited by R. O. Pohl
(Springer-Verlag, Berlin, 1993)p427;
 N. Nishiguchi, S. Tamura and F. Nori, Phys. Rev. B {\bf 48}, (1993) in press.
\bibitem{garg} A. Garg and D. Levine, Phys. Rev. Lett.
{\bf 60}, 2160 (1988); {\it ibid} {\bf 63}, 1439 (1989).
\bibitem{stoch1} P. Hoel, S. Port  and C. Stone, {\it Introduction to
Stochastic Processes} (Houghton, Boston, 1972)
\bibitem{stoch2} N.G. Van Kampen, {\it Stochastic Processes in Physics and
Chemistry} (North-Holland, Amsterdam, 1981)
\bibitem{qc} D. Levine and P. J. Steinhardt, Phys. Rev. B {\bf 34}, 596
(1986).
\bibitem{CSM} Z. Cheng, R. Savit and R. Merlin,  Phys. Rev. B {\bf 37}, 4375
   (1988).
\bibitem{tn90} S. Tamura and F. Nori, Phys. Rev. B {\bf 41}, 7941 (1990).
\bibitem{abrikosov}
A.~A.~Abrikosov and I.~A.~Ryzhkin, Adv. Phys. {\bf 27}, 147 (1978).
\bibitem{lfs} See, for example, I. M. Lifshitz, S. A. Postur and E. Yankovsky,
{\it Introduction to the Theory of Disordered Systems} (Wiley, New York,1988).
\bibitem{abrssc}
A.~A.~Abrikosov,   Solid State Commun. {\bf 37}, 997 (1981).
\bibitem{andrsn} P.W.~Anderson, D.J.~Thouless, E.~Abrahams and D.S.~Fisher,
Phys. Rev. B {\bf 22}, 3519(1980).
\bibitem{Ishii} K. Ishii, Prog. Theor. Phys. Suppl.  {\bf 53}, 77(1973).
\bibitem{Kono} K. Kono and S. Nakada, Phys. Rev. Lett.  {\bf 69},
  1185(1992).

\end{references}
\end{document}